# Easy-Setup Eye Movement Recording System for Human-Computer Interaction

Manh Duong PHUNG, Quang Vinh TRAN
Department of Electronics and Computer Engineering
College of Technology, Vietnam National University
Hanoi, Vietnam

Kenji HARA, Hirohito INAGAKI, Masanobu ABE
NTT Cyber Solutions Laboratories
NTT Corporation
Yokosuka, Japan

*Abstract*— Tracking the movement of human eyes is expected to yield natural and convenient applications based on human-computer interaction (HCI). To implement an effective eye-tracking system, eye movements must be recorded without placing any restriction on the user's behavior or user discomfort. This paper describes an eye movement recording system that offers free-head, simple configuration. It does not require the user to wear anything on her head, and she can move her head freely. Instead of using a computer, the system uses a visual digital signal processor (DSP) camera to detect the position of eye corner, the center of pupil and then calculate the eye movement. Evaluation tests show that the sampling rate of the system can be 300 Hz and the accuracy is about 1.8 °/s.

## I. Introduction

The movement of human eyes contains valuable information that can enhance the interaction between people and computers. For example, in a feedback system, users can select an answer by fixating it for a certain amount of time without operating mouse or keyboard. This reduces the time taken for selection and brings to users a more natural interaction. Another example appears in human face to face communication. While talking to each other, movements of the eyes play an important role in regulating the conversation. It is promising to integrate the analysis of eye movements into a modern man machine interface [1].

A number of techniques now have been proposed to capture eye movements, such as electro-oculography (EOG), the sclera search coil method (SSC), the photoelectric method, and videooculography (VOG) [2,3]. Each methodology has its strengths and limitations. The EOG technique measures signal intensity obtained from periphery muscles associated with eye movements. The photoelectric technique tracks the limbus of the eye by measuring the amount of scattered light. Both EOG and photoelectric methods suffer from a number of limitations including unreliably vertical eye movement measurements, low sensitivity and limited bandwidth [4,5,6]. The sclera search coil technique is generally considered the most reliable method as it offers the best spatial and temporal resolution, without limiting the range of eye movements. However, this method requires the subject to place coils onto the conjunctiva which usually produces some discomfort after about 30 min of recording [7]. In recent years, with the improvements in computing power and image processing, the VOG is becoming more and more attractive, due to its high accuracy, insignificant artifacts and wide applicability [2,8]. At present, most VOG systems use infra-red cameras to get high contract eye images; the user's eyes are illuminated by an infra-red source. Research shows that this may result in potential eye hazards such as eye dryness, lens burning, and retina thermal injury [9,10]. In addition, systems normally require the user to wear additional devices on her head. This will disturb the user and it is a bother to put on and take off even light head devices.

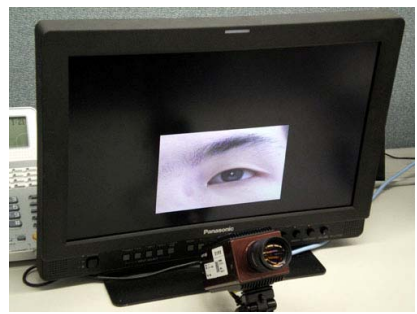

Figure 1. Overview of the eye movement recording system

In this paper, we propose a VOG-based eye movement recording system that employs only a visual DSP camera; head movement is free and no addition devices need be worn. An overview of the system is given in fig.1. The camera captures images at the spatial resolution of 640 pixels by 480 pixels. The pupil area is covered partially by an eyelid and eyelashes. In this study, we suppose that the iris and pupil are concentric circles.

## II. Eye Movement Recording Method

The proposed system records eye movements by two processes. First, the center of pupil is detected from the captured image by a double circle fitting algorithm. The variance projection function (VPF) is then employed to detect the eye corner. Let's say $p_1$, $c_1$ are alternatively the center of pupil and the eye corner at time $t_1$. $p_2$, $c_2$ are alternatively the center of pupil and the eye corner at time $t_2$. The average velocity of eye movement between $t_1$ and $t_2$ can be computed

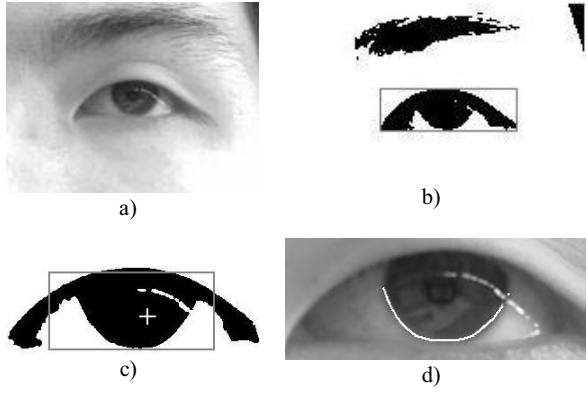

a)                  b)

c)                  d)

Figure 2. Sample detection processes

a) Captured eye image, b) Image segmentation and detected eye area, c) Refined eye area and detected iris position, d) Detected sample points.

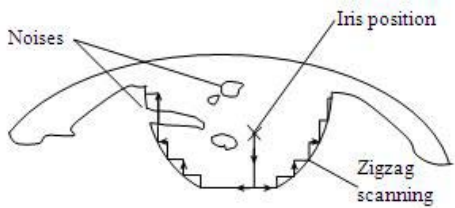

Figure 3. Detecting sample points by zigzag scanning

by:

$$v = \frac{(p_2 - c_2) - (p_1 - c_1)}{t_2 - t_1} \quad (1)$$

### A. Pupil center detection

The center of the pupil is detected by the logical steps described below.

First, an eye image is captured at a resolution of 640x480 [Fig.2 (a)]. The image is assumed to consist of the eyebrow and eye regions. This simplifies eye detection which is investigated in previous research. To reduce the processing cost, an 80x60 resolution image is generated from the original image. A threshold value is dynamically calculated using isodata algorithm [11]. A segmented image is generated by connecting neighborhood pixels [Fig.2 (b)].

Based on this segmented image, the eyebrow and eye regions are extracted as two largest regions and the eye area is the region below the eyebrow [Fig.2 (b)]. It is then copied to a buffer. All other regions are ignored.

A window is used to detect the iris position by scanning the eye region horizontally. Its height is the height of eye region and its width is 0.15 times the eye region width. These values are set based on the ratio of iris to eye dimensions. During the scanning process, the sum of grey-scale values inside the window is calculated. The position at which the window reaches its maximum value is the position of iris [Fig.2 (c)].

Using the iris position, the eye area is refined and the threshold value is recalculated. This allows the system to work in different lighting conditions.

Now, we return to the original image. All known data included threshold value, eye region and iris position are rescaled. To detect the sample points, points in the iris but at the iris-sclera border, we check the following conditions:

- Starting from the iris position, a vertical scan is made, using the grey-scale threshold, to detect the iris pixel just inside the iris-sclera border.
- The scan is then extended left and right by raising the scan line height by one pixel and extending it outward until the next sample point is discovered.

With these steps, the system detects sample points by following zigzag paths instead of line by line as conventional. As the result, the detecting process is fast and noise inside or at the edge of the iris is ignored [Fig.3]. Extracted samples are input to the circle fitting algorithm [Fig.2(d)].

The circle fitting algorithm is based on minimizing the mean square distances from the circle to the sample points [12]. Given n points $(x_i, y_i)$, $1 \leq i \leq n$, the objective function is defined by:

$$F(a, b, R) = \sum_{i=1}^{n} (\sqrt{(x_i - a)^2 + (y_i - b)^2} - R)^2 \quad (2)$$

where $(a, b)$ is the center of circle and R is its radius. The problem is: determine $(a, b, R)$ to F minimized.

There is no direct algorithm for computing the minimum of F, all known algorithms are either iterative (geometric fit) or approximate (algebraic fit) by nature. In this paper, we choose algebraic fit because of its high performance.

In algebraic fit, instead of minimizing the sum of squares of the geometric distances, we minimize the sum of squares of algebraic distances. F becomes

$$F(a, b, R) = \sum_{i=1}^{n} (z_i + Bx_i + Cy_i + D)^2 \quad (3)$$

Where:

$z_i = x_i^2 + y_i^2, B = -2a, C = -2b, D = a^2 + b^2 - R^2$

Differentiating F with respect B, C, D yields a system of linear equations:

$$M_{xx}B + M_{xy}C + M_x D = -M_{xz}$$
$$M_{xy}B + M_{yy}C + M_y D = -M_{yz} \quad (4)$$
$$M_x B + M_y C + nD = -M_z$$

where $M_{xx}$, $M_{xy}$, etc. denote moments, for example $M_{xx} = \sum x_i^2$, $M_{xy} = \sum x_i y_i$. Solving this system by Cholesky decomposition gives B, C, D and finally we compute (a, b, R).

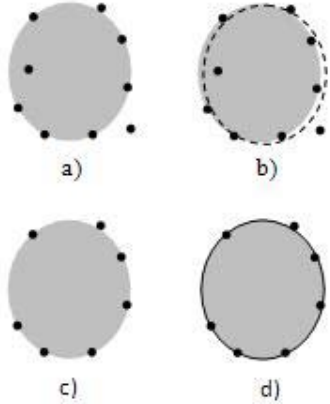

Figure 4. Double circle fitting

a) Sample points, b) first fit of circle,
c) Noise removing, d) Estimated circle.

As explained in fig.4, after estimating (a, b, R), the distances between sample points and the center of circle are calculated. Sample points which are far from the center are considered noise and are eliminated. Again, circle fitting is performed and the final center is extracted. We call this method *double circle fitting*. Fig.6 shows typical circles estimated by our method.

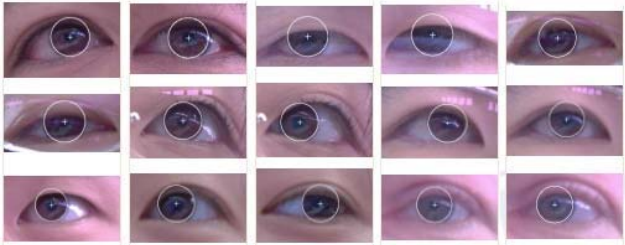

Figure 5. Center pupil detection

### B. Eye corner detection

In eye-based interacting systems, the influence of head displacement on eye movement determination is a problem that needs solving. Previous research eliminated this influence by requiring the user to follow an additional process such as calibration or putting a marker on the user's face [8,13]. However, any additional process is time consuming and bothers the user. In our system, we use an eye corner as a reference point to automatically offset head movement.

To detect the eye corner, a variance project function (VPF) is employed [14]. This method is based on the observation that some eye landmarks such as eye corners have relatively high contrast which can be detected effectively by the VPF. Suppose

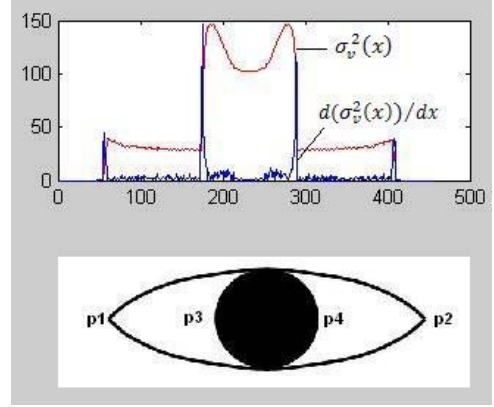

Figure 6. A synthetic image and its VPF along the vertical direction

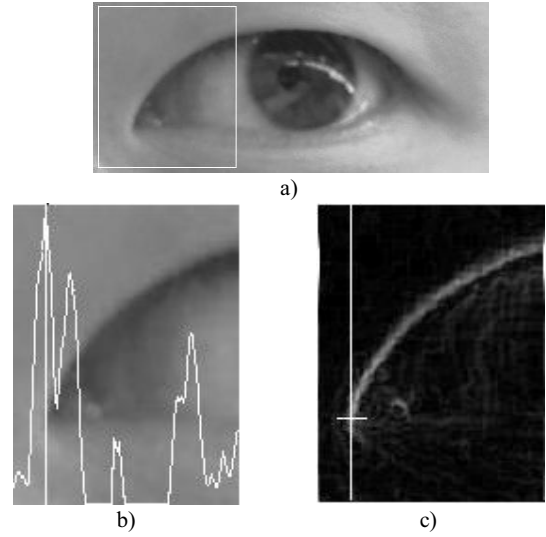

Figure 7. Eye corner detection

a) Eye corner area, b) Derivative of VPF, c) Eyelid and detected eye corner

$I(x, y)$ is the intensity of a pixel at location $(x, y)$, the variance projection function in vertical direction $\sigma_v^2(x)$ in interval $[y_1, y_2]$ is defined as follow:

$$\sigma_v^2(x) = \frac{1}{y_2 - y_1} \sum_{y_i = y_1}^{y_2} [I(x, y_i) - V_m(x)]^2 \quad (5)$$

where, $V_m(x)$ is mean of the vertical integral projection of $I(x, y)$:

$$V_m(x) = \frac{1}{y_2 - y_1} \int_{y_1}^{y_2} I(x, y) dy \quad (6)$$

Fig.6 presents the VPF along the vertical direction of an synthetic eye. It suggests the following eye corner detection approach.

First, an eye corner area is determined based on the center of pupil and the radius of iris [Fig.7(a)]. The vertical variance projection function $\sigma_v^2(x)$ and its derivative are then applied to the area to detect the vertical position of the eye corner [Fig.7(b)]. A Sobel edge detector is used to detect the eyelid. The horizontal position is the intersection between the vertical position and the eyelid [Fig.7(c)].

## III. EVALUATION

We implemented an eye movement recording system that detects eye movements in real time. An overview of the system is shown in Fig.1. It has a programmable camera and a display. There is no computer in the system. The camera is equipped with a digital signal processor TMS320DM642 600MHz of Texas Instrument, an 128Mbyte SDRAM, an 1Mbyte flash memory and some other extension boards. The eye captured image has a resolution of 640x480 pixels and the sampling rate is 30 fps (frame per second).

To evaluate the accuracy and the performance of the system, a number of experiments were conducted.

### A. Subjects

Participants were eight men, two women. During the experiments, participants were required to keep their eyes 70cm from the display and were asked to fix their heads at 0°.

### B. Procedure

- **Fixation tests:** Subjects were told to move their eyes between two points at $\pm 16°$ in synchronization with an audio signal that had a constant beat. After moving to a point, the eyes were fixed at the point for 2s.

- **Smooth pursuit tests:** Subjects were told to follow a red point that moved horizontally on the screen with sinusoidal velocity. The maximum velocity of the sinusoid was 20 °/s and the period was 2s. The experiment is then repeated with the vertical movement of the red point.

- **Range tests:** A horizontal array of points was displayed on the display. Eye movement from one point to another corresponds to an eye displacement of about 3°. Subject traced out a pattern, starting at the center point, and then moving to next outward point. The fixation time at each point was approximately 3s. Subjects were then required to repeat the experiment with a vertical array.

- **Performance tests:** During operating process of the system, the algorithm was repeatedly applied to each image frame until the system was suspended.

### C. Results

Fig.9 plots the results of the fixation tests. It shows that the average velocity of the horizontal eye movement was 1.12°/s, for the vertical eye movement it was 2.48°/s. If we assume that the subjects fixed their eyes at each point during the fixation time, these velocities can be considered to indicate the accuracy of the system.

Fig.10 shows the results gained when following sinusoidal movements. It shows that it is difficult for the subjects to follow a steady movement smoothly. This matches the inherent behavior of human eyes which emphasizes fixation and saccadic movements.

The linear ranges of our system are $\pm 19.65°$ in x-direction and $\pm 15.94°$ in y-direction.

At the 30 fps (frame per second), the system works normally when the algorithm is applied 10 times to each image frame. This means that the system can work at 300 Hz sampling rate.

### D. Interaction with survey machine

To demonstrate the eye movement recording system, an eye movement interaction application called survey machine was developed [Fig.8]. At the beginning of the demonstration, the system starts automatically when it detects user's eyes for 3s. An agent then appears and poses questions to the user. The user can select the answer by looking at Yes or No button on the screen which corresponds to left or right eye movement. Ten people interacted with the system.

During the interacting process, eye detection was sometimes failed and user had to try again. The reason is that user's hair covered the eyebrow result in wrong detection of eye area which was not focused by our system.

After the demonstration, all subjects enjoyed this new interaction and asked about the system. In fact, applications of the system are not limited to surveys; the many application areas include games, device controlling and information board.

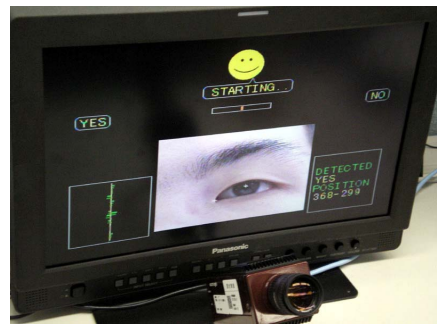

Figure 8. Survey machine

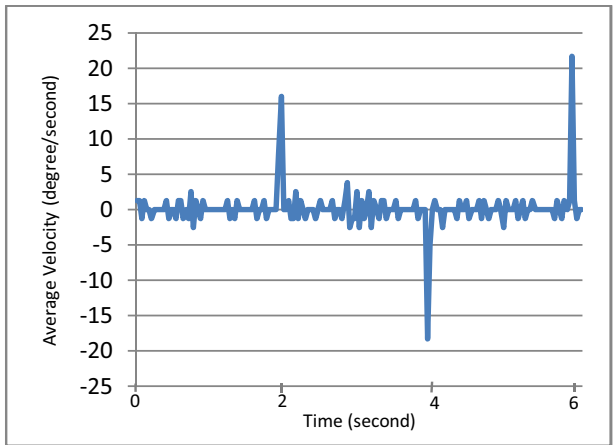
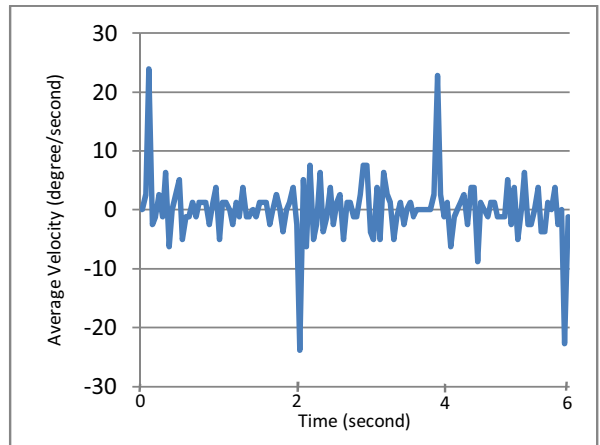

Figure 9. Results of fixation tests
a) Horizontal movement; b) Vertical movement

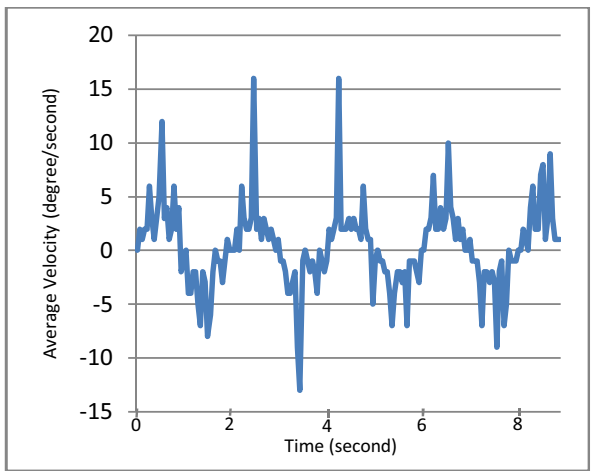
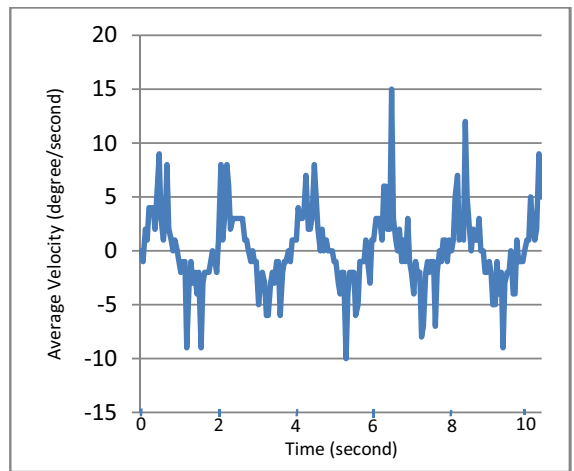

Figure 10. Results of smooth pursuit tests
a) Horizontal movement; b) Vertical movement

IV. DISCUSSION

After analyzing the results of the experiments, we found that the system successfully recorded the movement of the user's eyes without requiring any additional devices to be worn. The average error rate was $1.80^0$/s. The error in the Y direction is higher than the X direction because the upper and lower parts of the eye are covered by the eyelids. To reduce this error, a compensative technique needs to be developed.

During the experiment, the system sometimes failed to detect the eye movements. The reason may be that a large part of the iris is hidden which resulted in insufficient number of sample points [Fig.11]. It is difficult to detect eye movement in these situations.

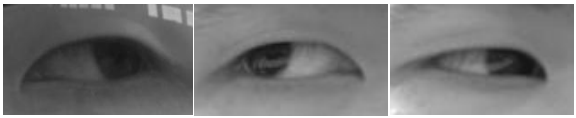

Figure 11. Failed detection situations

V. CONCLUSION AND FUTURE WORK

An easy-setup eye movement recording system was developed for human-computer interaction. It uses robust pupil center and eye corner detection algorithms. A number of evaluation tests and a demonstration were undertaken by ten people; the results prove the accuracy and the applicability of the system.

In future work, we will improve the accuracy of the system and a high speed camera will be used. We also have a plan to apply the eye movement recording system to game applications.

ACKNOWLEDGMENT

This research was supported by NTT Cyber Solutions Laboratories, NTT corporation. We are grateful to the members of Network Appliance and Services Project for their invaluable comments and advice on this research.